\documentclass[superscriptaddress,prl,twocolumn,amsmath,amssymb]{revtex4-1}
\pdfoutput=1
\usepackage{graphicx}
\usepackage{eurosym}
\usepackage{amsmath}
\usepackage{amssymb}
\usepackage{dcolumn}
\usepackage{bm}
\usepackage{subfigure}
\usepackage[final]{pdfpages}
\usepackage{soul}
\usepackage{comment}
\usepackage{mathtools}

\def\avg#1{\mathinner{\langle{#1}\rangle}}
\def\bra#1{\mathinner{\langle{#1}|}}
\def\ket#1{\mathinner{|{#1}\rangle}}
\newcommand{\braket}[2]{\langle #1|#2\rangle}

\newcommand{\ignore}[1]{}

\newcommand{\be}{\begin{equation}}
\newcommand{\ee}{\end{equation}}
\newcommand{\ba}{\begin{eqnarray}}
\newcommand{\ea}{\end{eqnarray}}

\DeclarePairedDelimiter\ceil{\lceil}{\rceil}

\begin{document}

\title{Exploiting locality in quantum computation for quantum chemistry}
\date{\today}

\author{Jarrod R. McClean} 
\affiliation{Department of Chemistry and Chemical Biology, Harvard University, Cambridge, MA 02138}
\author{Ryan Babbush} 
\affiliation{Department of Chemistry and Chemical Biology, Harvard University, Cambridge, MA 02138}
\author{Peter J. Love} 
\affiliation{Department of Physics, Haverford College, Haverford, PA 19041}
\author{Al\'an Aspuru-Guzik}
\affiliation{Department of Chemistry and Chemical Biology, Harvard University, Cambridge, MA 02138}

\begin{abstract} 
Accurate prediction of chemical and material properties from first principles quantum chemistry is a challenging task on traditional computers.  Recent developments in quantum computation offer a route towards highly accurate solutions with polynomial cost, however this solution still carries a large overhead.  In this perspective, we aim to bring together known results about the locality of physical interactions from quantum chemistry with ideas from quantum computation. We show that the utilization of spatial locality  combined with the Bravyi-Kitaev transformation offers an improvement in the scaling of known quantum algorithms for quantum chemistry and provide numerical examples to help illustrate this point. We combine these developments to improve the outlook for the future of quantum chemistry on quantum computers.
\end{abstract}

\maketitle

\section{Introduction}
Within chemistry, the Schr\"odinger equation encodes all information required to predict chemical properties ranging from reactivity in catalysis to light absorption in photovoltaics.  Unfortunately the exact solution of the Schr\"odinger equation is thought to require exponential resources on a classical computer, due to the exponential growth of the dimensionality of the Hilbert space as a function of molecular size.  This makes exact methods intractable for more than a few heavy atoms~\cite{Thogersen2004}.  

Richard Feynman first suggested that this scaling problem might be overcome if a more natural approach was taken~\cite{Feynman1982}.  Specifically, instead of painstakingly encoding quantum information into a classical computer, one may be able to use a quantum system to naturally represent another quantum system and bypass the seemingly insurmountable storage requirements.  This idea eventually developed into the field of quantum computation, which is now believed to hold promise for the solution of problems ranging from factoring numbers~\cite{Shor:1994tb} to image recognition~\cite{Neven:2008} and protein folding \cite{Perdomo-Ortiz2012,Babbush2012}.

Initial studies by Aspuru-Guzik et. al. showed that these approaches might be particularly promising for quantum chemistry~\cite{Aspuru:2005}.  There have been many developments both in theory~\cite{Jones:2012,Seeley2012,Yung:2013} and experimental realization~\cite{Lanyon:2010,Walther:2012,Peruzzo2013} of quantum chemistry on quantum computers.  The original gate construction for quantum chemistry introduced by Whitfield et al.~\cite{Whitfield:2011} was recently challenged as too expensive by Wecker et al.~\cite{Wecker:2013}. The pessimistic assessment was due mostly to the extrapolation of the Trotter error for artificial rather than realistic molecular systems, as was analyzed in detail in a followup study by many of the same authors~\cite{Poulin:2014}. They subsequently improved the scaling by means of several circuit enhancements~\cite{Hastings:2014}. 
The analysis of the Trotter error on realistic molecules in combination with their improvements led to a recent study where an estimate of the calculation time of Fe$_2$S$_2$ was reduced by orders of magnitude~\cite{Poulin:2014}.
In this paper, we further reduce the scaling by exploiting the locality of physical interactions with local basis sets as has been done routinely now in quantum chemistry for two decades~\cite{Car:1985,Goedecker:1999}.  These improvements in combination with others make quantum chemistry on a quantum computer a very attractive application for early quantum devices.  We describe the scaling under two prominent measurement strategies, quantum phase estimation and Hamiltonian averaging, which is a simple subroutine of the recently introduced Variational Quantum Eigensolver approach~\cite{Peruzzo2013}.

Additionally, recent progress in accurate and scalable solutions of the Schr\"odinger equation on classical computers has also been significant~\cite{Car:1985,Artacho:1999,Goedecker:1999,Helgaker2002,Shao:2006,Bowler:2012}.  Some of these results have already appeared in the quantum computation literature in the context of in depth studies of state preparation~\cite{Wang:2008,Veis:2014}.  A general review of quantum simulation~\cite{Buluta:2009,Georgescu:2014} and one on quantum computation for chemistry~\cite{Kassal:2011} cover these topics in more depth.  A collection covering several aspects of quantum information and chemistry recently appeared~\cite{Kais:2014}. However many developments that utilize fundamental physical properties of the systems being studied to enable scalability have not yet been exploited.  

In this study, we hope to bring to light results from quantum chemistry as well as their scalable implementation on quantum computers.  We begin by reviewing the standard electronic structure problem.  Results based on the locality of physical interactions from linear scaling methods in quantum chemistry are then introduced with numerical studies to provide quantification of these effects.  A discussion of the resulting impact on the most common quantum algorithms for quantum chemistry follows. We also investigate instances where a perfect oracle is not available to provide input states, demonstrating the need for advances in state preparation technology. Finally, we conclude with an outlook for the future of quantum chemistry on quantum computers.

\section{Electronic structure problem}
To frame the problem and set the notation, we first briefly introduce the electronic structure problem of quantum chemistry~\cite{Helgaker2002}.  Given a set of nuclei with associated charges $\{Z_i\}$ and a total charge (determining the number of electrons), the physical states of the system can be completely characterized by the eigenstates $\{\ket{\Psi_i}\}$ and corresponding eigenvalues (energies) $\{E_i\}$ of the Hamiltonian $H$
\begin{align}
 H &= - \sum_i \frac{\nabla_{R_i}^2}{2 M_i} - \sum_i \frac{\nabla_{r_i}^2}{2} - \sum_{i,j} \frac{Z_i}{|R_i - r_j|} \notag \\
 &+ \sum_{i, j>i} \frac{Z_i Z_j}{|R_i - R_j|} + \sum_{i, j>i} \frac{1}{|r_i - r_j|}
\end{align}
where we have used atomic units, $\{R_i\}$ denote nuclear coordinates, $\{r_i\}$ electronic coordinates, $\{Z_i\}$ nuclear charges, and $\{M_i\}$ nuclear masses. Owing to the large difference in masses between the electrons and nuclei, typically the Born-Oppenheimer approximation is used to mitigate computational cost and the nuclei are treated as stationary, classical point charges with fixed positions $\{R_i\}$.  Within this framework, the parametric dependence of the eigenvalues on $\{R_i\}$, denoted by $\{E(\{R_i\})_j\}$ determines almost all chemical properties, such as bond strengths, reactivity, vibrational frequencies, etc.  Work has been done in the determination of these physical properties directly on a quantum computer~\cite{Kassal:2009}.

Due to the large energy gaps between electronic levels with respect to the thermal energy scale $k_B T$, it typically suffices to study a small subset of the eigenstates corresponding to the lowest energies.  Moreover, for this reason, in many molecules the lowest energy eigenstate $\ket{\Psi_0}$, or ground state, is of primary importance, and for that reason it is the focus of many methods, including those discussed here.

\subsection{Second quantized Hamiltonian}
Direct computation in a positional basis accounting for anti-symmetry in the wavefunction while using the Hamiltonian described is referred to as a first quantization approach and has been explored in the context of quantum computation~\cite{Kassal:2008,Ward:2009,Welch:2014}.  The first quantized approach has also been realized in experiment~\cite{Lu:2011}. One may also perform first quantized calculations in a basis of slater determinants. This was introduced as a representation of the electronic wavefunction by qubits in ~\cite{Aspuru:2005} (the compact mapping) and the efficiency of time evolution in this basis was recently shown~\cite{Toloui:2013}. The second quantized approach places the antisymmetry requirements on the operators.   After choosing some orthogonal spin-orbital basis $\{\varphi_i\}$ with a number of terms $M$, the second quantized Hamiltonian may be written as ~\cite{Helgaker2002}
\begin{align}
 \hat H = \sum_{pq} h_{pq} a_p^{\dagger}a_q + \frac{1}{2} \sum_{pqrs} h_{pqrs} a_p^{\dagger} a_q^{\dagger} a_r a_s
\end{align}
with coefficients determined by
\begin{align}
 &h_{pq} = \int d\sigma\ \varphi_p^*(\sigma) \left(-\frac{\nabla_{r}^2}{2} - \sum_{i} \frac{Z_i}{|R_i - r|} \right)\varphi_q(\sigma) \\
 &h_{pqrs} = \int d\sigma_1\ d\sigma_2\ \frac{ \varphi_p^*(\sigma_1) \varphi_q^*(\sigma_2)  \varphi_r(\sigma_1) \varphi_s(\sigma_2) }{|r_1 - r_2|}
\end{align}
where $\sigma_i$ now contains the spatial and spin components of the electron, $\sigma_i = (r_i, s_i)$.  The operators $a_p^\dagger$ and $a_r$ obey the fermionic anti-commutation relations
\begin{align}
 \{a_p^\dagger, a_r\} &= \delta_{p,r} \\
 \{a_p^\dagger, a_r^\dagger\} &= \{a_p, a_r\} = 0
\end{align}
For clarity, we note that the basis functions used in quantum chemistry (such as atom-centered Gaussians) are frequently parameterized on the nuclear coordinates $\{R_i\}$, which can result in a dependence on the nuclear positions of the electronic integral terms $\{h_{pqrs}\}$. For notational simplicity the dependence of the integrals on the nuclear positions in this work will remain implied.

\subsection{Spatial locality}
It is clear by inspection that the maximum number of terms in the second-quantized Hamiltonian scales as $O(M^4)$.  $M$ can be quite large to reach chemical accuracy for systems of interest, and the number of terms present in the Hamiltonian is a dominant cost factor for almost all quantum computation algorithms for chemistry.  However, due to the locality of physical interactions, one might imagine that many of the terms in the Hamiltonian are negligible relative to some finite precision $\epsilon$.   
While this depends on the basis, it is this observation that forms the foundation for the linear-scaling methods of electronic structure such as linear scaling density functional theory or quantum Monte Carlo~\cite{Artacho:1999,Williamson:2001,Aspuru:2005}.  That is, in a local basis, the number of non-negligible terms scales more like $O(M^2)$, and advanced techniques such as fast multipole methods techniques can evaluate their contribution in $O(M)$ time.  

These scaling properties are common knowledge within the domain of traditional quantum chemistry, however they have not yet been exploited within the context of quantum computation.  They are clearly vitally important for the correct estimate of the asymptotic scaling of any method~\cite{Aspuru:2005,Whitfield:2011,Jones:2012,Wecker:2013}.  For that reason, we review the origin of that scaling here for the most common and readily available local basis, the Gaussian atomic orbital basis.  We follow loosely the explanation presented by Helgaker, J{\o}rgensen, and Olsen~\cite{Helgaker2002}, and refer readers to this text for additional detail on the evaluation of molecular integrals in local basis sets.  The two elements we will consider here are the cutoffs due to exponentially vanishing overlaps between Gaussians basis functions and a bound on the value of the largest integral.

By far the most common basis used in electronic structure calculations is a set of atom-centered Gaussian (either Cartesian or ``Pure'' spherical) functions.  While the precise result can depend on the angular momentum associated with the basis function, for simplicity, consider only Gaussian $S$ functions, which is defined by
\begin{align}
 \ket{G_a} = \exp \left( -a r_A^2 \right)
\end{align}
where $r_A$ is the vector from a point $A$ which defines the center of the Gaussian. One property of Gaussian functions that turns out to be useful in the evaluation of molecular integrals is the Gaussian product rule.  This rule states simply that the product of two spherical Gaussian functions may be written in terms of a single spherical Gaussian function on the line segment connecting the two centers.  Consider two spherical Gaussian functions, $\ket{G_a}$ and $\ket{G_b}$ separated along the $x$-axis.
\begin{align}
 \exp \left( -a x_A^2 \right) \exp \left( -b x_B^2 \right) 
  = K^x_{ab} \exp \left( -p x_p^2 \right)
\end{align}
where $K^x_{ab}$ is now a constant pre-exponential factor
\begin{align}
 K^x_{ab} = \exp \left(-\mu X^2_{AB} \right)
\end{align}
and the total exponent $p$, the reduced exponent $\mu$, and the Gaussian separation $X_{AB}$ are given by
\begin{align}
 p &= a + b \\
 \mu &= \frac{ab}{a+b} \\
 X_{AB} &= A_x - B_x
\end{align}
That is, the product of two spherical Gaussians is a third Gaussian centered between the original two that decays faster than the original two functions, as given by the total exponent $p$.
The overlap integral of two spherical Gaussian $S$ functions may be obtained through application of the Gaussian product rule after factorizing into the three Cartesian dimensions, followed by Gaussian integration and is given by
\begin{align}
 S_{ab} = \braket{G_a}{G_b} = \left( \frac{\pi}{a + b} \right)^{3/2} \exp \left( - \frac{ab}{a + b} R^{2}_{AB} \right)
\end{align}
where $R_{AB}$ is the distance between the Gaussian centers $A$ and $B$.  Clearly this integral decays exponentially with the square of the distance between centers, and one may determine a distance $d_s$ such that beyond that distance, the integrals will be smaller than $10^{-k}$ in magnitude.
\begin{align}
 d_s = \sqrt{ a_{\text{min}}^{-1} \log \left[ \left( \frac{\pi}{2 a_{\text{min}}}\right)^3 10^{2k} \right] }
\end{align}
where $a_{\text{min}}$ is the minimal Gaussian exponent $a$ (most diffuse function) in the set of Gaussian basis functions $\{\ket{G_a}\}$. While the exact decay parameters will depend on the basis set, it is generally true from this line of reasoning that there is a characteristic distance, beyond which all overlap integrals are negligible. This means that the number of interactions per basis function becomes fixed, resulting in a linear number of significant overlap integrals. As kinetic energy integrals are just fixed linear combinations of overlap integrals of higher angular momentum, the same argument holds for them as well.

For $S$ orbitals, the two-electron Coulomb integral may be written as
\begin{align}
h_{acbd} = \frac{S_{ab}S_{cd}}{R_{PQ}} \text{\ erf}(\sqrt{\alpha} R_{PQ})
\end{align}
where erf is the error function, $P$ and $Q$ are Gaussian centers formed through application of the Gaussian product rule to $\ket{G_a} \ket{G_b}$ and $\ket{G_c}\ket{G_d}$ respectively.  $R_{PQ}$ is the distance between the two Gaussian centers $P$ and $Q$ and $\alpha$ is the reduced exponent derived from $P$ and $Q$.  For clarity, this may be bounded by the simpler expression
\begin{align}
h_{acbd} \leq \min \left( \frac{4 \alpha}{\pi} S_{ab} S_{cd}, \frac{S_{ab} S_{cd}}{R_{PQ}} \right)
\end{align}
The first of these two expressions in the min function comes from the short range bound and the latter from the long range bound of the error function.  These bounds show that the integrals are determined by products of overlap terms, such that in the regime where overlap integrals scale linearly, we expect $O(M^2)$ significant two-electron terms.  Moreover, as seen in the long range bound of the two-electron integral, there is some further asymptotic distance beyond which these interactions may be completely neglected, yielding an effectively linear scaling number of significant integrals.  This limit can be quite large however, thus practically one expects to observe a quadratic scaling in the number of two-electron integrals (TEI).

Additionally, we note from the form of the integrals, that the maximal values the two-electron integrals will attain are determined by the basis set parameters, such as the width of the Gaussian basis functions or their angular momentum.  The implication of this, is that the maximal integral magnitude for the four index two-electron integrals, $|h^{\text{TEI}}_{\text{max}}|$ will be independent of the molecular size for standard atom centered Gaussian basis sets, and may be treated as a constant for scaling analysis that examine cost as a function of physical system size with fixed chemical composition.  The overlap and kinetic energy integrals will similarly have a maximum independent of molecular size past a very small length scale.  However, the nuclear attraction integrals must also be considered.

While not typically considered a primary source of difficulty due to the relative ease of evaluation with respect to two-electron integrals, we separate the nuclear attraction integrals here due to the fact that the maximal norm of the elements may change as well.  The nuclear attraction matrix element between $S$ functions may be written as
\begin{align}
 h_{ab}^{\text{nuc}} = -\sum_{i} \frac{Z_i S_{ab}}{R_{Pi}}\text{\ erf}\left(\sqrt{p} R_{Pi} \right)
\end{align}
where $Z_i$ is the nuclear charge and $R_{Pi}$ refers to the distance between the Gaussian center $P$ with total exponent $p$ formed from the product $\ket{G_a}\ket{G_b}$ to the position of the $i$'th nuclei.  Following from the logic above, from the exponentially vanishing overlap $S_{ab}$, at some distance, we expect only a linear number of these integrals to be significant.  However, each of the integrals considers the sum over all nuclei, which can be related linearly to the number of basis functions in atom centered Gaussian basis sets.  Thus the maximal one-electron integral is not a constant, but rather can be expected to scale with the Coulomb sum over distant nuclear charges.  A conservative bound can be placed on such a maximal element as follows.  

The temperature and pressure a molecule reside in will typically determine the minimal allowed separation of two distinct nuclei, and will thus define a maximum nuclear density $\rho_{\text{max}}$.  Denote the maximum nuclear charge in the systems under consideration as $Z_{\text{max}}$.  The maximal density and the number of nuclei will also define a minimal radius that a sphere of charge may occupy $r_{\text{max}}$,
\begin{align}
 r_{\text{max}}^3 = \frac{3 Z_{\text{max}} N_{\text{nuc}} }{4 \pi \rho_{\text{max}}}
\end{align}
where $N_{\text{nuc}}$ is the number of nuclei in the system. Modeling the charge as spread uniformly within this minimal volume and using the maximum of the error function to find a bound on the maximum for the nuclear attraction matrix element, we find
\begin{align}
 |h_{ab}^{\text{nuc}}| &< 4 \pi \rho_{\text{max}} S_{ab} \left| \int_{0}^{r_{\text{max}}} \ r^2 dr \frac{1}{r}\right| \notag \\
 &=  2 \pi \rho_{\text{max}} S_{ab} r_{\text{max}}^2 \notag \\
 &= \beta_{ab} N_{\text{nuc}}^{2/3}
\end{align}
where $\beta_{ab}$ is now a system size independent quantity determined only by basis set parameters at nuclei $a$ and $b$, and the size dependence is bounded as $O(N_{\text{nuc}}^{2/3})$.  Atom centered Gaussian basis sets will have a number of a basis functions which is a linear multiple of the number of nuclei, and as such we may now bound the maximal one-electron integral (OEI) element as
\begin{align}
 |h_{\text{max}}^{\text{OEI}}| < \beta_{\text{max}}^{\text{OEI}} M^{2/3}
\end{align}

\subsection{Effect of truncation}
The above analysis demonstrates that given some integral magnitude threshold, $\delta$, there exists a characteristic distance $d$ between atomic centers, beyond which integrals may be neglected.  If one is interested in a total precision $\epsilon$ in the energy $E_i$, it is important to know how choosing $\delta$ will impact the solution, and what choice of $\delta$ allows one to retain a precision $\epsilon$.

By specification, the discarded integrals are small with respect to the rest of the Hamiltonian (sometimes as much as $10$ orders of magnitude smaller in standard calculations).  As such, one expects a perturbation analysis to be accurate.  Consider the new, truncated Hamiltonian $H_t = H + V$, where $V$ is the negation of the sum of all removed terms, each of which have magnitude less than $\delta$.

Assuming a non-degenerate spectrum for $H$, from perturbation theory we expect the leading order change in eigenvalue $E_i$ to be given by
\begin{equation}
 \Delta E_i = \bra{\Psi_i} V \ket{\Psi_i}
\end{equation}
if the number of terms removed from the sum is given by $N_r$, a worst case bound on the magnitude of this deviation follows from the spectrum of the creation and annihilation operators and is given by
\begin{equation}
 |\Delta E_i| \leq \sum_{\{h_i : |h_i| < \delta \}} |h_{i}| \leq N_r \delta
\end{equation}
where ${\{h_i : |h_i| < \delta \}}$ is simply the set of Hamiltonian elements with norm less than $\delta$ and the first inequality follows directly from the triangle inequality. We emphasize that this is a worst case bound, and generically one expects at least some cancellation between terms, such as kinetic and potential terms, when the Hamiltonian is considered as a whole.  Some numerical studies of these cancellation effects have been performed~\cite{Poulin:2014}, but additional studies are required.  Regardless, under this maximal error assumption, by choosing a value 
\begin{equation} 
  \delta \leq \frac{\epsilon}{N_r}
\end{equation}
one retains an accuracy $\epsilon$ in the final answer with respect to the exact answer when measuring the eigenvalue of the truncated Hamiltonian $H_t$.  Alternative, one may use the tighter bound based on the triangle inequality and remove the maximum number of elements such that the total magnitude of removed terms is less than $\epsilon$.  From the looser but simpler bound, we see a reduction of scaling from $M^4$ to $M^2$ would require removal of the order of $M^4$ terms from the Hamiltonian, this constraint on $\delta$ can be rewritten in terms of $M$ as
 \begin{equation} 
  \delta \leq \frac{\epsilon}{M^4}
\end{equation}

While the perturbation of the eigenvalue will have a direct influence on energy projective measurement methods such as quantum phase estimation, other methods evaluate the energy by averaging.  In this case, we do not need to appeal to perturbation theory, and the $\delta$ required to achieve a desired $\epsilon$ can be found directly.
\begin{align}
 \langle H_t \rangle &= \bra{\Psi_i} H_t \ket{\Psi_i} \\
 &= E_i + \bra{\Psi_i} V \ket{\Psi_i}
\end{align}
We find that under our assumption of worst case error for averaging, the result is identical to that of the first order perturbation of the eigenvalue $E_i$, 
\begin{align}
|\Delta \langle H_t \rangle| \leq \sum_{\{h_i : |h_i| < \delta \}} |h_{i}| \leq N_r \delta
\end{align}
In summary, we find that for both the consideration of the ground state eigenvalue and the average energy of the ground state eigenvector, there is a simple formula for the value of $\delta$, which scales polynomially in the system size, below which one may safely truncate to be guaranteed an accuracy $\epsilon$ in the final answer.  Moreover it suggests a simple strategy that one may utilize to achieve the desired accuracy.  That is, sort the integrals in order of magnitude, and remove the maximum number of integrals such that the total magnitude of removed integrals is less than $\epsilon$.

On the subject of general truncation, we note that while there exist may Hamiltonians with the same structure as the second quantized electronic structure Hamiltonian that have the property that removal of small elements will cause a drastic shift in the character of the ground state, this has not been seen for physical systems in quantum chemistry.  In practice it is observed that removing elements on the order of $\delta = 10^{-10}$ and smaller is more than sufficient to retain both qualitative and quantitative accuracy in systems of many atoms~\cite{Artacho:1999,Williamson:2001,Helgaker2002,Aspuru:2005}.  Moreover, the convergence with respect to this value may be tested easily for any systems under consideration.

\subsection{Onset of favorable scaling}
While the above analysis shows that locality of interactions  in local basis sets provides a promise that beyond a certain length scale, the number of non-negligible integrals will scale quadratically in the number of basis functions, it does not provide good intuition for the size of that length scale in physical systems of interest.  Here we provide numerical examples for chemical systems in basis sets used so far in quantum computation for quantum chemistry.  The precise distance at which locality starts to reduce the number of significant integrals depends, of course, on the physical system and the basis set used.  In particular, larger, more diffuse basis sets are known to exhibit these effects at comparatively larger length scales than minimal, compact basis sets.  However the general scaling arguments given above hold for all systems of sufficient size.
\begin{figure}
\centering
\includegraphics[width=8cm]{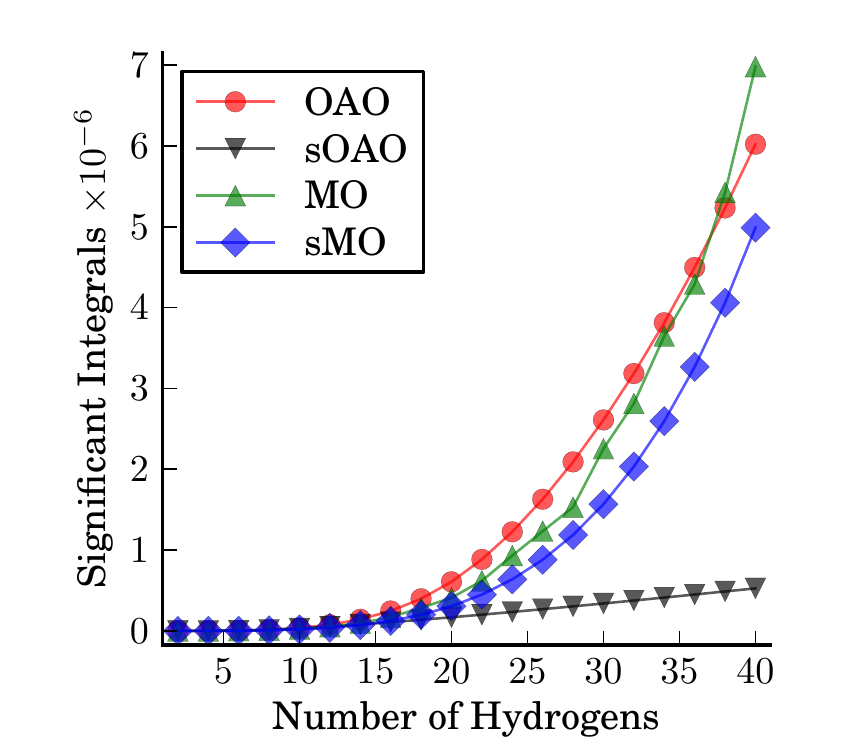}
\caption{The number of significant(magnitude $> 10^{-15}$) spin-orbital integrals in the STO-3G basis set as a function of the number of hydrogens in a linear hydrogen chain with a separation of 1 $a_0$ for the Hartree-Fock canonical molecular orbital basis(MO) and the symmetrically orthogonalized atomic orbital basis (OAO).  The sMO and sOAO, shows the same quantity with a sharper cutoff ($10^{-7}$) and demonstrates the advantage to localized atomic basis functions at length scales as small as 10 \AA.}
\label{fig:hydrogenPlot}
\end{figure}

An additional consideration which must be made for quantum computation, is that as of yet, no general technology has been developed for direct simulation in non-orthogonal basis sets.  This prohibits direct simulation in the bare atomic orbital basis, however the use of L\"owdin symmetric orthogonalization yields the orthogonal basis set closest to the original atomic orbital basis set in an $l^2$ sense~\cite{Lowdin:1950,Mayer:2002}.  We find that this is sufficient for the systems we consider, but note that there have been a number of advances in orthogonal basis sets that are local in both the occupied and virtual spaces and may find utility in quantum computation ~\cite{Ziolkowski:2009}. Moreover, there has been recent work in the use of multiresolution wavelet basis sets that have natural sparsity and orthogonality while providing provable error bounds on the choice of basis~\cite{Harrison:2004}.  Such a basis also allows one to avoid costly integral transformations related to orthogonality, which are known to scale as $O(M^5)$ when performed exactly.  Further research is needed to explore the implications for quantum computation with these basis sets, and we focus here on the more common atom-centered Gaussian basis sets.

As a prototype system, we consider chains of hydrogen atoms separated by 1 Bohr ($a_0$) in the STO-3G basis set, an artificial system that can exhibit a transition to a strongly correlated wavefunction~\cite{Hachmann:2006}.  We count the total number of significant integrals for values of $\delta$ given by $10^{-15}$ and $10^{-7}$ for the symmetrically orthogonalized atomic orbital (OAO) basis and the canonical Hartree-Fock molecular orbital (MO) basis.  The results are displayed in Fig. \ref{fig:hydrogenPlot} and demonstrate that with a cutoff of $\delta=10^{-7}$ the localized character of the OAO's allows for a savings of on the order of $6\times 10^{6}$ integrals with respect to the more delocalized canonical molecular orbitals.   The s in the labeling of the orbital bases simply differentiates between two possible cutoffs.  These dramatic differences begin to present with atomic chains as small as 10 \AA \ in length in this system with this basis set.

\begin{figure}
\centering
\includegraphics[width=8cm]{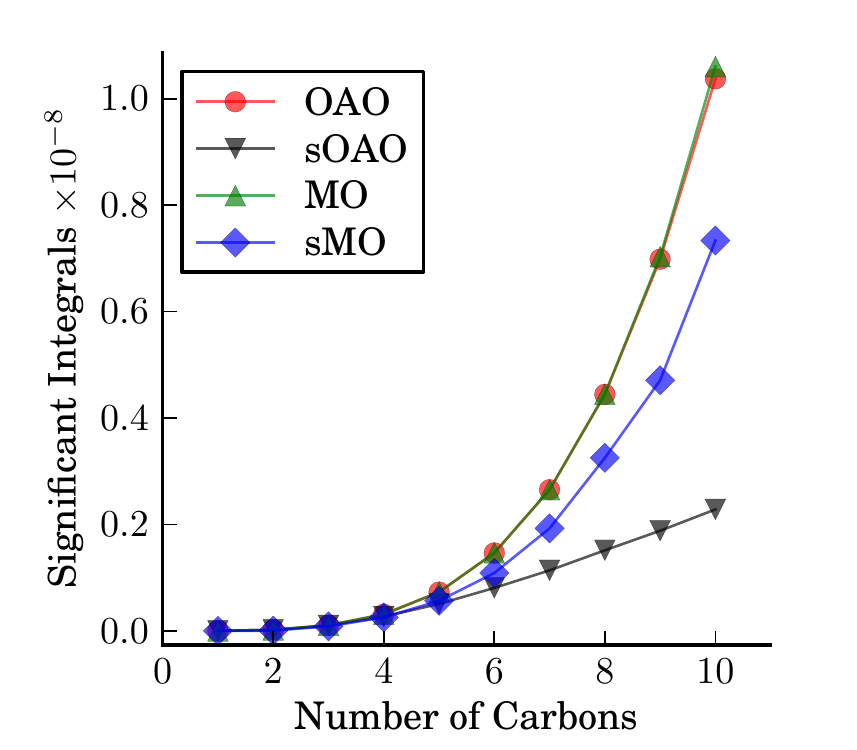}
\caption{The number of significant(magnitude $> 10^{-15}$) spin-orbital integrals in the STO-3G basis set as a function of the number of carbons in a linear alkane chain for the Hartree-Fock canonical molecular orbital basis(MO) and the symmetrically orthogonalized atomic orbital basis (OAO).  The sMO and sOAO shows the same quantity with a sharper cutoff ($10^{-7}$) and demonstrates the dramatic advantage to localized atomic basis even at this small atomic size.}
\label{fig:alkanePlot}
\end{figure}

As an additional example, we consider linear alkane chains of increasing length.  The results are displayed in Fig. \ref{fig:alkanePlot} and again display the dramatic advantages of preserving locality in the basis set.  By the point one reaches 10 carbon atoms, a savings of almost $10^{8}$ integrals can be achieved at a truncation level of $10^{-7}$.

Although localized basis sets provide a definitive scaling advantage in the medium-large size limit for molecules, one often finds that in the small molecule limit canonical molecular orbitals, the orbitals from the solution of the Hartree-Fock equations under the canonical condition, provide a more sparse representation.  This is observed in Figs \ref{fig:hydrogenPlot} and \ref{fig:alkanePlot} for the smallest molecule sizes, and the transition for this behavior will generally be basis set dependent.  The reason is that at smaller length scales, the ``delocalized'' canonical molecule orbitals have similar size to the more localized atomic orbitals, but with the additional constraint of the canonical condition, a sufficient but not necessary condition for the solution of the Hartree-Fock equations that demands the Fock matrix be diagonal (as opposed to the looser variational condition of block-diagonal between the occupied and virtual spaces).  A side effect of the canonical condition is that in the canonical molecular orbital basis many of the $h_{pqrs}$ terms for distinct indices are reduced in magnitude.  However, there are not enough degrees of freedom present in the orbital rotations for this effect to persist to larger length scales, and as a result local basis sets eventually become more advantageous.  Moreover, it is known that at larger length scales, the canonical conditions tend to favor maximally delocalized orbitals, which can reduce the advantages of locality.  These effects have been studied in some detail in the context of better orbital localizations by relaxing the canonical condition in Hartree-Fock and the so-called Least-Change Hartree-Fock method coupled with fourth-moment minimization~\cite{Ziolkowski:2009}.

\section{Quantum energy estimation}
Almost all algorithms designed for the study of quantum chemistry eigenstates on a quantum computer can be separated into two distinct parts: 1. state preparation and 2. energy estimation.  For the purposes of analysis, it is helpful to treat the two issues separately, and in this section we make the standard assumption in doing so, that an oracle capable of producing good approximations to the desired eigenstates $\ket{\Psi_i}$ at unit cost is available.  Under this assumption, energy estimation for a fixed desired precision $\epsilon$ is known to scale polynomially in the size of the system for quantum chemistry, however the exact scaling costs and tradeoffs depend on the details of the method used.  Here we compare the costs and benefits of two prominent methods of energy estimation used in quantum computation for chemistry: quantum phase estimation and Hamiltonian averaging.

\subsection{Quantum phase estimation}
The first method used for the energy estimation of quantum chemical states on a quantum computer was quantum phase estimation~\cite{Kitaev:1995,Abrams1999,Aspuru:2005}.  The method works by evolving the give quantum eigenstate $\ket{\Psi_i}$  forward under the system Hamiltonian $H$ for a time $T$, and reading out the accumulated phase, which can be easily mapped to the associated eigenenergy $E_i$.  While the basic algorithm and its variations can have many different components, the cost is universally dominated by the coherent evolution of the system.

To evolve the system under the Hamiltonian, one must find a scalable way to implement the unitary operator $U=e^{-i H T}$.  The standard procedure for accomplishing this task is the use of Suzuki-Trotter splitting~\cite{Trotter:1959,Suzuki:1993}, which approximates the unitary operator(at first order) as
\begin{align}
 U&=e^{-i H T} = \left(e^{-i H (T/m)}\right)^m \notag \\ 
 &= \left(e^{-i \left( \sum_i H_i \right) \Delta t}\right)^m
 \approx \left(\prod_i e^{-i H_i \Delta t} \right)^m
\end{align}
where $\Delta t = T/m$ and $H_i$ is a single term from the Bravyi-Kitaev transformed system Hamiltonian. Higher order Suzuki-Trotter operator splittings and their benefits have been studied in the context of quantum simulation~\cite{Berry:2007}, but we largely focus on the first order formula in this work. If each of the simpler unitary operators $e^{-i H_i \Delta t}$ has a known gate decomposition, the total time evolution can be performed by chaining these sequences together.  

The use of the Suzuki-Trotter splitting can be thought of as an evolution under an approximate Hamiltonian $\tilde H$, given by $e^{-i \tilde H T}$, whose eigenspectrum deviates from the original Hamiltonian by a factor depending on time-step $\Delta t$.  The precise dependence of this bias depends on the order of the Suzuki-Trotter expansion used. The total resolution, $\epsilon$, in the energies of the approximate Hamiltonian $\tilde H$ is determined by the total evolution time $T$.  Thus to achieve an accuracy of $\epsilon$ in the final energy, one must utilize a time step $\Delta t$ small enough that the total bias is less than $\epsilon$ and a total run time $T$ such that the resolution is better than $\epsilon$. If the number of gates required to implement a single timestep $\Delta t$ is given by $N_g$, then the dominant cost of simulation (all of which must be done coherently) is given by
\begin{equation}
 N_c = N_g \ceil*{\frac{T}{\Delta t}}
\end{equation}

The total evolution time $T$ required to extract an eigenvalue to chemical precision $\epsilon_{\text{chem}}=10^{-3}$ is typically set at the Fourier limit independent of molecular size and thus can be considered a constant for scaling analysis.  We then focus on the number of gates per Suzuki-Trotter time step, $N_g$, and the time step $\Delta t$ required to achieve the desired precision.

In a first order Suzuki-Trotter splitting, the number of gates per Trotter time step is given by the number of terms in the Hamiltonian multiplied by the number of gates required to implement a single elementary term for the form $e^{-i H_i \Delta t}$.  The gates per elementary term can vary based on the particular integral, however for simplicity in developing bounds we consider this as constant here.  The number of terms, is known from previous analysis in this work to scale as $O(M^2)$ or in the truly macroscopic limit $O(M)$.  The number of gates required to implement a single elementary term depends on the transformation used from fermionic to qubit operators.  The Jordan-Wigner transformation~\cite{Jordan1928} results in non-local terms that carry with them an overhead that scales as the number of qubits, which in this case will be $O(M)$.  Although there have been developments in methods to use teleportation to perform these non-local operations in parallel~\cite{Jones:2012} and by improving the efficiency of the circuits computing the phases in the Jordan-Wigner transformation~\cite{Hastings:2014}, these issues can also be alleviated by choosing the Brayvi-Kitaev transformation that carries an overhead only logarithmic in the number of qubits, $O(\log M)$~\cite{Bravyi:2002,Seeley2012}.  As a result, one expects the number of gates per Suzuki-Trotter time step $N_g$ to scale as $O(M^2 \log M)$ or in a truly macroscopic limit $O(M \log M)$. 

To complete the cost estimate with fixed total time $T$, one must determine how the required time step $\Delta t$ scales with the size of the system.  As mentioned above, the use of the Suzuki-Trotter decomposition for the time evolution of $H$ is equivalent to evolution under an effective Hamiltonian $\tilde H = H+V$, where the size of the perturbation is determined by the order of the Suzuki-Trotter formula used and the size of the timestep.  Once the order of the Suzuki-Trotter expansion to be used has been determined, the requirement on the timestep is such that the effect of $V$ on the eigenvalue of interest is less than the desired accuracy in the final answer $\epsilon$. 

This has been explored previously~\cite{Hastings:2014,Poulin:2014}, but we now examine this scaling in our context.  To find $V$, one may expand the $k$'th order Suzuki-Trotter expansion of the evolution of $\tilde H$ into a power series as well as the power series of the evolution operator $\exp \left[-i \left( H + V \right) \Delta t \right]$, and find the leading order term $V$.  As a first result, we demonstrate that for a $k$'th order propagator, the leading perturbation on the ground state eigenvalue for a non-degenerate system is $O(\Delta t)^{k+1}$.

Recall the power series expansion for the propagator 
\begin{align}
\exp \left[-i \left( H + V \right) \Delta t \right] = \sum_{j=0}^{\infty} \frac{(-i)^j}{j!} \left(H + V\right)^j \left(\Delta t \right)^j
\end{align}
The definition of a $k$'th order propagator, is one is that correct through order $k$ in the power series expansion.  As such, when this power series is expanded, $V$ must make no contribution in the terms until $O( (\Delta t)^{k+1})$.  For this to be possible, it's clear that $V$ must depend on $\Delta t$.  In order for it to vanish for the first $k$ terms, $V$ must be proportional to $(\Delta t)^{k}$.  Moreover, due to the alternation of terms between imaginary and real at each order in the power series with the first term being imaginary, the first possible contribution is order $(\Delta t)^{k}$ and imaginary.  As is common in quantum chemistry, we assume a non-degenerate and real ground state, and thus the contribution to the ground state eigenvalue is well approximated by first order perturbation theory as
\begin{align}
 E^{(1)} = \bra{\Psi_g} V \ket{\Psi_g}
\end{align}
however, as $V$ is imaginary Hermitian and the ground state is known to be real in quantum chemistry, this expectation value must vanish.  Thus the leading order perturbation to the ground state eigenvalue is at worst the real term depending on $(\Delta t)^{k+1}$.

To get a more precise representation of $V$ for a concrete example, we now consider the first order $(k=1)$ Suzuki-Trotter expansion.  As expected, the leading order imaginary error term is found to be
\begin{align}
 V^{(0)} = \frac{\Delta t}{2} \sum_{j < k} i \left[H_j, H_k \right] 
\end{align}
whose contribution must vanish due to it being an imaginary Hermitian term.  Thus we look to the leading contributing error depending on $(\Delta t)^{2}$, which has been obtained previously\cite{Poulin:2014} from a Baker-Campbell-Hausdorff(BCH) expansion to read
\begin{align}
 V^{(1)} &= \frac{(\Delta t)^2}{12} \sum_{i \leq j} \sum_j \sum_{k < j} \left[ H_i\left(1 - \frac{\delta_{ij}}{2}\right), \left[H_j, H_k\right] \right]
\end{align}
Thus the leading order perturbation is given by third powers of the Hamiltonian operators.  To proceed, we count the number of one- and two-electron integrals separately as $N_{\text{int}}^{\text{OEI}}$ and $N_{\text{int}}^{\text{TEI}}$ respectively.  Their maximal norm elements are similarly denoted by $h_{\text{max}}^{\text{OEI}}$ and $h_{\text{max}}^{\text{TEI}}$.  From this, we can draw a worst case error bound on the perturbation of the eigenvalue given by 
\begin{align}
 E^{(1)} &\leq \frac{(\Delta t)^2}{12} \sum_{i \leq j} \sum_j \sum_{k < j} \left| H_i\left(1 - \frac{\delta_{ij}}{2}\right), \left[H_j, H_k\right] \right|  \notag\\ 
  &\leq \left(|h_{\text{max}}^{\text{OEI}}|N_{\text{int}}^{OEI} + |h_{\text{max}}^{\text{TEI}}| N_{\text{int}}^{\text{TEI}}\right)^3 \left(\Delta t\right)^2 \\
  &\leq \left(|\beta_{\text{max}}^{\text{OEI}}|M^{2/3}N_{\text{int}}^{\text{OEI}} + |h_{\text{max}}^{\text{TEI}}| N_{\text{int}}^{\text{TEI}}\right)^3 \left(\Delta t\right)^2 \notag
\end{align}
Where the first inequality follows from the triangle inequality and the second is a looser, but simpler bound, that may be used to elucidate the scaling behavior. 
Holding the looser bound to the desired precision in the final answer $\epsilon$, this yields
\begin{align}
 \Delta t \leq \left[\frac{\epsilon}{\left(|\beta_{\text{max}}^{\text{OEI}}|M^{2/3}N_{\text{int}}^{\text{OEI}} + |h_{\text{max}}^{\text{TEI}}| N_{\text{int}}^{\text{TEI}}\right)^3} \right]^{1/2}
\end{align}

We emphasize that this is a worst case bound, including no possible cancellation between Hamiltonian terms.  Some preliminary work has been done numerically in establishing average cancellation between terms that shows these worst case bounds are too pessimistic~\cite{Poulin:2014}.  Continuing, we expect the total scaling under a first order Suzuki-Trotter expansion using a Bravyi-Kitaev encoding to be bounded by
\begin{align}
 N_c &= N_g \ceil*{\frac{T}{\Delta t}} \leq \frac{N_g }{\epsilon \Delta t} \notag \\
     &\leq \frac{\left(|\beta_{\text{max}}^{\text{OEI}}|M^{2/3}N_{\text{int}}^{\text{OEI}} + |h_{\text{max}}^{\text{TEI}}| N_{\text{int}}^{\text{TEI}}\right)^{3/2} N_{\text{int}} \log M}{\epsilon^{3/2}}
\end{align}
and in the large size limit where the number of significant two-electron integrals in a local basis set scales quadratically and the number of significant one-electron integrals scales linearly, this may be bounded by
\begin{align}
 N_c \leq \kappa \frac{\left(|\beta_{\text{max}}^{\text{OEI}}|M^{5/3} + |h_{\text{max}}^{\text{TEI}}| M^2\right)^{3/2}(M^2 + M) \log M}{\epsilon^{3/2}}
\end{align}
where $\kappa$ is a positive constant that will depend on the basis set and this expression scales as $O(M^5 \log M)$ in the number of spin-orbital basis functions.
\subsection{Hamiltonian averaging}
The quantum phase estimation algorithm has been central in almost all algorithms for energy estimation in quantum simulation.  However, it has a significant practical drawback in that after state preparation, all the desired operations must be performed coherently.  A different algorithm for energy estimation has recently been introduced~\cite{Peruzzo2013,Yung:2013} that lifts all but an $O(1)$ coherence time requirement after state preparation, making it amenable to implementation on quantum devices in the near future.  We briefly review this approach, which we will call Hamiltonian averaging, and bound its costs in applications for quantum chemistry.

As in quantum phase estimation, in Hamiltonian averaging one assumes the eigenstates $\ket{\Psi_i}$ are provided by some oracle. By use of either the Jordan-Wigner or Bravyi-Kitaev transformation, the Hamiltonian may be written as a sum of tensor products of Pauli operators.  These transformations at worst conserve the number of independent terms in the Hamiltonian, thus we may assume for our worst case analysis the number of terms is fixed by $N_{int}$ and the coefficients remain unchanged.  From the provided copy of the state and transformed Hamiltonian, to obtain the energy one simply performs the average
\begin{align}
 \langle \hat H \rangle = \sum_{i,j,k,... \in {x,y,z}} h_{ijk...} \langle \sigma_1^i \otimes \sigma_2^j \otimes \sigma_3^k... \rangle
\end{align}
by independent Pauli measurements on the provided state $\ket{\Psi_i}$ weighted by the coefficients $h_{ijkl...}$, which are simply a relabeling of the previous two-electron integrals for convenience with the transformed operators.  As $\ket{\Psi_i}$ is an eigenstate, this average will correspond to the desired eigenvalue $E_i$ with some error related to sampling that we now quantify.

Consider an individual term
\begin{align}
 X_{ijkl...} = h_{ijkl...} \sigma_1^i \otimes \sigma_2^j \otimes \sigma_3^k...
\end{align}
it is clear from the properties of qubit measurements, that the full range of values this quantity can take on is $[-h_{ijkl...}, h_{ijkl...}]$.  As a result, we expect that the variance associated with this term can be bounded by
\begin{align}
 \text{Var} \left[ X_{ijkl...} \right] \leq |h_{ijkl...}|^2
\end{align}

Considering a representative element, namely the maximum magnitude integral element $h_{\text{max}}$, we can bound the variance of $\hat H $ as
\begin{align}
\text{Var} \left[ \hat H \right] \leq N_{int}^2 |h_{\text{max}}|^2
\end{align}

The variance of the mean, which is the relevant term for our sampling error, comes from the central limit theorem and is bounded by
\begin{align}
  \text{Var} \left[ \langle \hat H \rangle \right] \leq \frac{\text{Var} \left[ \hat H \right]}{N}
\end{align}
where $N$ is the number of independent samples taken of $\avg{\hat H}$. Collecting these results, we find
\begin{align}
  \text{Var} \left[ \langle \hat H \rangle \right] &\leq \sum \frac{|h_{ijkl...}|^2}{N} \notag \\
   &\leq \frac{\left(|\beta_{\text{max}}^{\text{OEI}}|M^{2/3}N_{\text{int}}^{\text{OEI}} + |h_{\text{max}}^{\text{TEI}}| N_{\text{int}}^{\text{TEI}}\right)^{2}}{N}
\end{align}
Now setting the variance to the desired statistical accuracy $\epsilon^2$ (which corresponds to a standard error of $\epsilon$ at a 68\% confidence interval), we find the number of independent samples expected, $N_s$, is bounded by
\begin{align}
 N_s \leq \frac{\left(|\beta_{\text{max}}^{\text{OEI}}|M^{2/3}N_{\text{int}}^{\text{OEI}} + |h_{\text{max}}^{\text{TEI}}| N_{\text{int}}^{\text{TEI}}\right)^{2}}{\epsilon^2}
\end{align}
If a single independent sample of $\avg{\hat H}$ requires the measurement of each of the $N_{int}$ quantities, then the bound on the total cost in the number of state preparations and measurements, $N_m$ is
\begin{align}
 N_{m} \leq \frac{N_{int} \left(|\beta_{\text{max}}^{\text{OEI}}|M^{2/3}N_{\text{int}}^{\text{OEI}} + |h_{\text{max}}^{\text{TEI}}| N_{\text{int}}^{\text{TEI}}\right)^{2} }{\epsilon^2}
\end{align}
which if one considers the large size limit,such that the number of two-electron integrals scales quadratically and the number of one-electron integrals scales linearly, we find
\begin{align}
 N_{m} \leq \kappa \frac{(M+M^2) \left(|\beta_{\text{max}}^{\text{OEI}}|M^{5/3}  + |h_{\text{max}}^{\text{TEI}}| M^2 \right)^{2} }{\epsilon^2}
\end{align}
where $\kappa$ is a positive constant that depends upon the basis set.  It is clear that this expression scales as $O(M^6)$ in the number of spin-orbital basis functions.  We see from this, that under the same maximum error assumptions, Hamiltonian averaging scales only marginally worse in the number of integrals and precision as compared to quantum phase estimation performed with a first order Suzuki-Trotter expansion, but has a coherence time requirement of $O(1)$ after each state preparation.  Note that each measurement is expected to require single qubit rotations that scale as either $O(M)$ for the Jordan-Wigner transformation or $O(\log M)$ for the Bravyi-Kitaev transformation.  However, we assume that these trivial single qubit rotations can be performed in parallel independent of the size of the system without great difficulty, and we thus don't consider this in our cost estimate. This method is a suitable replacement for quantum phase estimation in situations where coherence time resources are limited and good approximations to the eigenstates are readily available.  Additional studies are needed to quantify the precise performance of the two methods beyond worst case bounds.

\section{Using imperfect oracles}
A central assumption for successful quantum phase estimation and typically any energy evaluation scheme is access to some oracle capable of producing good approximations to the eigenstate of interest, where a ``good'' approximation is typically meant to imply an overlap that is polynomial in the size of the system.  Additionally, a supposed benefit of phase estimation over Hamiltonian averaging is that given such a good (but not perfect) guess, by projective measurement in the energy basis, in principle one may avoid any bias in the final energy related to the initial state.  Here we examine this assumption in light of the Van-Vleck catastrophe~\cite{VanVleck:1936}, which we review below, and examine the consequences for measurements of the energy by QPE and Hamiltonian averaging.

The Van Vleck catastrophe~\cite{VanVleck:1936} refers to an expected exponential decline in the quality of trial wavefunctions, as measured by overlap with the true wavefunction of a system, as a function of size.  We study a simple case of the catastrophe here in order to frame the consequences for quantum computation. Consider a model quantum system consisting of a collection of $N$ non-interacting two level subsystems with subsystem Hamiltonians given by $H_i$. These subsystems have ground and excited eigenstates $\ket{\psi_g^i}$ and $\ket{\psi_e^i}$ with eigenenergies $E_g < E_e$, such that the total Hamiltonian is given by
\begin{align}
 H = \sum_i H_i
\end{align}
and eigenstates of the total Hamiltonian are formed from tensor products of the eigenstates of the subsystems.  As such the ground state of the full system is given by
\begin{align}
 \ket{\Psi_g} = \bigotimes_{i=0}^{N-1} \ket{\psi_g^i}
\end{align}

Now suppose we want to measure the ground state energy of the total system, but the oracle is only capable of producing trial states for each subsystem $\ket{\psi_t^i}$ such that $\braket{\psi_t^i}{\psi_g^i} = \Delta$, where $|\Delta| < 1$. The resulting trial state for the whole system is
\begin{align}
 \ket{\Psi_t} = \bigotimes_{i=0}^{N-1} \ket{\psi_t^i}
\end{align}
From normalization of the two level system, we may also write the trial state as
\begin{align}
\ket{\psi_t^i} = \Delta \ket{\psi_g^i} + e^{-i\theta}\sqrt{1 - \Delta^2} \ket{\psi_e^i}
\end{align}
where $\theta \in [0, 2 \pi)$.  Moreover, from knowledge of the gap, one can find the expected energy on each subsystem, which is given by
\begin{align}
 \bra{\psi_t^i}H_i\ket{\psi_t^i} = \Delta^2 E_g + (1-\Delta^2)E_e
\end{align}

For the case of Hamiltonian averaging on the total system, the expected answer is given by
\begin{align}
 E &= \bra{\Psi_t}H\ket{\Psi_t} \notag\\
   &= \sum_{i=0}^{N-1} \bra{\psi^i_t}H_i\ket{\psi^i_t} \notag\\
   &= N (\Delta^2 E_g + (1-\Delta^2)E_e)
\end{align}
which yields an energy bias from the true ground state, $\epsilon_b$, given by
\begin{align}
 \epsilon_b &= N (\Delta^2 E_g + (1-\Delta^2)E_e) - N E_g \notag \\
          &= N (1 - \Delta^2) (E_e - E_g) \notag \\
          &= N (1 - \Delta^2) \omega
\end{align}
where we denote the gap for each subsystem as $\omega = (E_e - E_g)$.  As such, it is clear that the resulting bias is only linear in the size of the total system $N$.

Quantum phase estimation promises to remove this bias by projecting into the exact ground state.  However, this occurs with a probability proportional to the square of the overlap of the input trial state with the target state.  In this example, this is given by
\begin{align}
 |\braket{\Psi_t}{\Psi_g}| = |\Delta|^{2N}
\end{align}
which is exponentially small in the size of the system.  That is, quantum phase estimation is capable of removing the bias exactly in this example non-interacting system, but at a cost which is exponential in the size of the system.  The expected cost of removing some portion of the bias may be calculated by considering the distribution of states and corresponding energies.  

Consider first the probability of measuring an energy with a bias of $\epsilon(M) = M (1 - \Delta^2) \omega$.  For this to happen, it is clear that exactly $M$ of the subsystems in the measured state are in the excited state.  It is clear that this is true for $\left(\begin{array}{c} N \\ M \end{array} \right)$ eigenstates, and the square of the overlap with such an eigenstate is $\left(\Delta^2\right)^{N-M} \left(1-\Delta^2\right)^{M}$ or 
\begin{align}
 P(\epsilon(M)) = \left(\begin{array}{c} N \\ M \end{array} \right) \left(\Delta^2\right)^{N-M} \left(1-\Delta^2\right)^{M}
\end{align}
which is clearly a binomial distribution.  As a result, in the large $N$ limit, this distribution is well approximated by a Gaussian and we may write
\begin{align}
 P(\epsilon(M)) &\approx \frac{1}{\sqrt{2 \pi \sigma^2}} \exp \left[ - \frac{1}{2} \left(\frac{M - \bar N}{\sigma} \right)^2 \right] \\
 \bar N &= N (1 - \Delta^2) \\
 \sigma^2 &= N \Delta^2 (1 - \Delta^2)
\end{align}
Bringing this together, we find that the probability of measuring a bias of less than $\epsilon(M)$ is given by
\begin{align}
 P(< \epsilon(M)) &= \frac{1}{\sqrt{2 \pi \sigma^2}} \int_{0}^{M} \ dM'\exp \left[ - \frac{1}{2} \left(\frac{M' - \bar N}{\sigma} \right)^2 \right] \notag \\
 &= \frac{1}{2} \left[\text{erf} \left(\frac{M - \bar N}{\sqrt{2 \sigma^2}} \right) + \text{erf} \left(\frac{\bar N}{\sqrt{2 \sigma^2}} \right)\right]
\end{align}
where erf is again the error function.
\begin{figure}
\centering
\includegraphics[width=8cm]{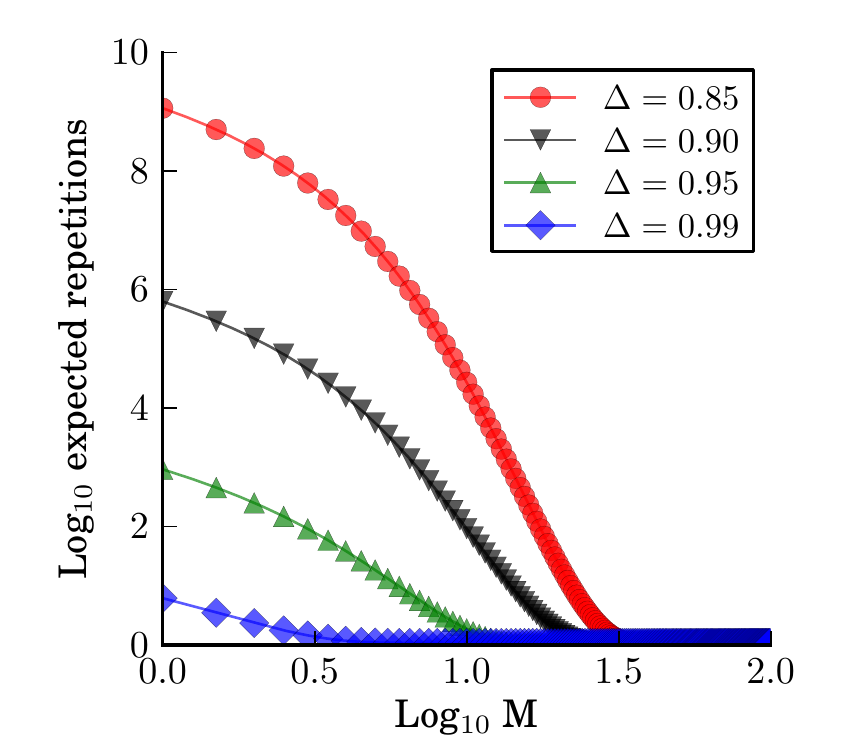}
\caption{A log log plot of the expected cost in number of repetitions of measuring an energy with a bias $\epsilon(M)$ as a function of $M$ in quantum phase estimation for different values of the oracle quality $\Delta$.  A system of $N=100$ non-interacting subsystems is considered.  A perfect, unbiased answer corresponds to $M=0$ with expected cost $O(\Delta^{2N})$, however to aid in visualization this plot is provided only beyond $M=1$.  In general one sees that depending on the oracle quality $\Delta$, different fractions of the bias may be removed with ease, but there is always some threshold for imperfect guesses ($|\Delta| < 1$) such that there is an exponential growth in cost.}
\label{fig:iErrorFunc}
\end{figure}

Thus the expected cost in terms of number of repetitions of the full phase estimation procedure to remove a bias of at least $\epsilon(M)$ from this model system is 
\begin{align}
 C(< \epsilon(M)) &= \frac{1}{P(< \epsilon(M))} \notag \\
  &= 2 \left[\text{erf} \left(\frac{M - \bar N}{\sqrt{2 \sigma^2}} \right) + \text{erf} \left(\frac{\bar N}{\sqrt{2 \sigma^2}} \right)\right]^{-1}
\end{align}
We plot the expected cost function for a range of oracle guess qualities $\Delta$ on a modest system of $N=100$ in Fig \ref{fig:iErrorFunc}. From this, we see that the amount of bias that can feasibly be removed depends strongly on the quality of the oracle guess.  Generically, we see that for any fixed imperfect guess on the subsystem level($|\Delta| < 1$), there will be an exponential cost in phase estimation related to perfect removal of the bias. 

This problem can be circumvented by improving the quality of the subsystem guesses as a function of system size.  In particular, one can see that if $|\Delta|$ is improved as $(1 - 1/(2N))$ then $|\Delta|^{2N}$ is $O(1)$.  However, as the subsystems in a general case could be of arbitrary size, classical determination of a subsystem state of sufficient quality may scale exponentially in the required precision and thus system size.  Moreover, one would not expect the problem to be easier in general cases where interactions between subsystems are allowed.  As a result, further developments in variational methods~\cite{Peruzzo2013}, quantum cooling~\cite{Xu:2014}, and adiabatic state preparation~\cite{Aspuru:2005,AQChem,Veis:2014} will be of key importance in this area.  Moreover improvements in the ansatze used to prepare the wave function such as multi-configurational self consistent field calculations(MCSCF)~\cite{Wang:2008,Veis:2014} or unitary coupled cluster (UCC)~\cite{Yung:2013} will be integral parts of any practical quantum computing for quantum chemistry effort.

\section{Adiabatic computation}
A complementary solution for the problem of molecular simulation on quantum computers is that of adiabatic quantum computation. It is not known to show the same direct dependence on the overlap of the initial guess state as QPE, which may allow it to solve different problems than the quantum phase estimation or variational quantum eigensolver in practice.  In \cite{AQChem}, Babbush \emph{et al.} show how to scalably embed the eigenspectra of molecular Hamiltonians in a programmable physical system so that the adiabatic algorithm can be applied directly. In this scheme, the molecular Hamiltonian is first written in second quantization using fermionic operators. This Hamiltonian is then mapped to a qubit Hamiltonian using the Bravyi-Kitaev transformation~\cite{Bravyi:2002,Seeley2012}. The authors show that the more typical Jordan-Wigner transformation cannot be used to scalably reduce molecular Hamiltonians to 2-local qubit interactions as the Jordan-Wigner transformation introduces linear locality overhead which translates to an exponential requirement in the precision of the couplings when perturbative gadgets are applied. Perturbative gadgets are used to reduce the Bravyi-Kitaev transformed Hamiltonian to a 2-local programmable system with a restricted set of physical couplings. Finally, tunneling spectroscopy of a probe qubit \cite{Berkley2013} can be used to measure eigenvalues of the prepared state directly.

While the exact length of time one must adiabatically evolve is generally unknown, Babbush \emph{et al.} argue that the excited state gap could shrink polynomially with the number of spin-orbitals when interpolating between exactly preparable noninteracting subsystems and the exact molecular Hamiltonian in which those subsystems interact. This would imply that adiabatic state preparation is efficient. Their argument is based on the observation that molecular systems are typically stable in their electronic ground states and the natural processes which produce these states should be efficient to simulate with a quantum device. Subsequently, Veis and Pittner analyzed adiabatic state preparation for a set of small chemical systems and observed that for all configurations of these systems, the minimum gap occurs at the very end of the evolution when the state preparation is initialized in an eigenstate given by a CAS (complete active space) ground state \cite{Veis:2014}. The notion that the minimum gap could be bounded by the physical HOMO (highest occupied molecular orbital) - LUMO (lowest unoccupied molecular orbital) gap lends support to the hypothesis put forward by Babbush \emph{et al.}

\subsection{Resources for adiabatic quantum chemistry}

In the adiabatic model of quantum computation, the structure of the final problem Hamiltonian (encoding the molecular eigenspectrum) determines experimental resource requirements. Since programmable many-body interactions are generally unavailable, we will assume that any experimentally viable problem Hamiltonian must be 2-local. Any 2-local Hamiltonian on $n$ qubits can be expressed as,
\begin{equation}
H = \alpha \cdot \textbf{1} + \sum_{i=1}^n \vec{\beta}_i \cdot \vec{\sigma}_i + \sum_{i=1}^{n-1} \sum_{j = i + 1}^n \vec{\gamma}_{ij}\cdot \left(\vec{\sigma}_i \otimes \vec{\sigma}_j\right)
\end{equation}
where $\vec{\sigma}_i = \left\langle \sigma_i^x, \sigma_i^y , \sigma_i^z \right\rangle$ is the vector of Pauli matrices on the $i^\textrm{th}$ qubit, $\alpha \in \mathbb{R}$ is a scalar and $\vec{\beta}_i \in \mathbb{R}^3$ and $\vec{\gamma}_{ij} \in \mathbb{R}^{9}$ are vectors of coefficients for each possible term.

In addition to the number of qubits, the most important resources are the number of qubit couplings and the range of field values needed to accurately implement the Hamiltonian. Since local fields are relatively straightforward to implement, we are concerned with the number of 2-local couplings,
\begin{equation}
\sum_{i=1}^{n-1} \sum_{j = i + 1}^n \textrm{card}\left(\vec{\gamma}_{ij}\right)
\end{equation}
where $\textrm{card}\left(\vec{v}\right)$ is the number of nonzero terms in vector $\vec{v}$. Since the effective molecular electronic structure Hamiltonian is realized perturbatively, there is a tradeoff between the error in the eigenspectrum of the effective Hamiltonian, $\epsilon$, and the strength of couplings that must be implemented experimentally. The magnitude of the perturbation is inversely related to the gadget spectral gap $\Delta$ which is directly proportional to the largest term in the Hamiltonian,
\begin{equation}
\max_{ij} \left\{\left\| \vec{\gamma_{ij}}\left(\epsilon\right)  \right\|_\infty\right\}\propto \Delta \left(\epsilon\right).
\end{equation}
Thus, the smaller $\Delta$ is, the easier the Hamiltonian is to implement but the greater the error in the effective Hamiltonian. In general, there are other important resource considerations but these are typically scale invariant; for instance, the geometric locality of a graph or the set of allowed interaction terms. The Hamiltonian can be modified to fit such constraints using additional perturbative gadgets but typically at the cost of using more ancilla qubits that require greater coupling strength magnitudes.

\subsection{Estimates of qubit and coupler scaling}
The Bravyi-Kitaev transformation is crucial when embedding molecular electronic structure in 2-local spin Hamiltonians due to the fact that this approach guarantees a logarithmic upper-bound on the locality of the Hamiltonian. A loose upper-bound (i.e. overestimation) for the number of qubits needed to gadgetize the molecular electronic Hamiltonian can be obtained by assuming that all terms have the maximum possible locality  of $O\left(\log\left(M\right)\right)$ where $M$ is the number of spin-orbitals.

In general, the number of terms produced by the Bravyi-Kitaev transformation scales the same as the number of integrals in the electronic structure problem, $O\left(M^4\right)$; however, as pointed out in an earlier section, this bound can be reduced to $O\left(M^2\right)$ if a local basis is used and small integrals are truncated. Using the ``bit-flip'' gadgets of \cite{Kempe2004, Jordan2008} to reduce $M^2$ terms of locality $\log\left(M\right)$, we would need $M^2 \log\left(M\right)$ ancillae. Since the number of ancilla qubits is always more than the number of logical qubits for this problem, an upper-bound on the total number of qubits needed is $O\left(M^2 \log\left(M\right)\right)$.

The number of couplings needed will be dominated by the number of edges introduced by ancilla systems required as penalty terms by the bit-flip gadgets. Each of the $O\left(M^2\right)$ terms is associated with a different ancilla system which contains a number of qubits equal to the locality of that term. Furthermore, all qubits within an ancilla system are fully connected. Thus, if we again assume that all terms have maximum locality, an upper-bound on the number of couplers is $O\left(M^2 \log^2\left(M\right)\right)$. Based on this analysis, the adiabatic approach to quantum chemistry has rather modest qubit and coupler requirements.

\subsection{Estimates of spectral gap scaling}
In \cite{AQChem}, Babbush \emph{et al.} reduce the locality of interaction terms using perturbative gadgets from the ``bit-flip'' family, first introduced in \cite{Kempe2004} and later generalized by \cite{Jordan2008}. In the supplementary material presented in a later paper analyzing the scaling of gadget constructions \cite{Cao2013}, it is shown that for bit-flip gadgets, $\lambda^{k+1}/\Delta^k = O\left(\epsilon\right)$ and
\begin{equation}
\max_{ij} \left\{\left\| \vec{\gamma_{ij}}\left(\epsilon\right)  \right\|_\infty\right\} = O\left(\frac{\lambda^k}{\Delta^{k-1}}\right).
\end{equation}
Here, $\lambda$ is the perturbative parameter, $\Delta$ is the spectral gap, $\epsilon$ is the error in the eigenspectrum and $\vec{\gamma}_{ij}$ is the coefficient of the term to be reduced. Putting this together and representing the largest coupler value as $\gamma$, we find that $\Delta = \Omega\left(\epsilon^{-k} \gamma^k\right)$, where $\Omega$ is the ``Big Omega'' lower bound. Due to the Bravyi-Kitaev transformation, the locality of terms is bounded by, $k = O\left(\log\left(M\right)\right)$; thus,  $\Delta = \Omega\left(\epsilon^{-\log\left(M\right)} \gamma^{\log\left(M\right)}\right)$. 

Prior analysis from this paper indicates that the maximum integral size is bounded by $\gamma \leq |\beta_{\text{max}}^{\text{OEI}}|M^{2/3}$. This gives us the bound,
\begin{equation}
\Delta = \Omega\left(\epsilon^{-\log\left(M\right)} \left\|\beta_{\text{max}}^{\text{OEI}}M^{2/3}\right\|^{\log\left(M\right)}\right).
\end{equation}
However, $\Delta$ also depends polynomially on $M^2$, the number of terms present. Though known to be polynomial, it is extremely difficult to predict exactly how $\Delta$ depends on $M^2$ as applying gadgets to terms ``in parallel'' leads to ``cross-gadget contamination'' which contributes at high orders in the perturbative expansion of the self-energy used to analyze these gadgets \cite{Cao2013}. Without a significantly deeper analysis, we can only conclude that,
\begin{equation}
\Delta = \Omega\left(\textrm{poly}\left(M\right)\left\|\frac{\beta_{\text{max}}^{\text{OEI}}M^{2/3}}{\epsilon}\right\|^{\log\left(M\right)}\right).
\end{equation}
This analysis indicates that the most significant challenge to implementing the adiabatic approach to quantum chemistry is the required range of coupler values which is certain to span \emph{at least} several orders of magnitude for non-trivial systems.

This calls attention to an important open question in the field of Hamiltonian gadgets: whether there exist ``exact'' gadgets which can embed the ground state energy of arbitrary many-body target Hamiltonians without the use of perturbation theory. A positive answer to this conjecture would allow us to embed molecular electronic structure Hamiltonians without needing large spectral gaps. For entirely diagonal Hamiltonians, such gadgets are well known in the literature \cite{Biamonte2008,Babbush2013b} but  fail when terms do not commute \cite{Cao2013}. Exact reductions have also been achieved for certain Hamiltonians. For instance, ``frustration-free'' gadgets have been used in proofs of the QMA-Completeness of quantum satisfiability, and in restricting the necessary terms for embedding quantum circuits in Local Hamiltonian~\cite{Nagaj2010,Gosset2013,Childs2013}.

\section{Conclusions}
In this work, we analyzed the impact on scaling for quantum chemistry on a quantum computer that results from consideration of locality of interactions and exploitation of local basis sets.  The impact of locality has been exploited to great advantage for some time in traditional algorithms for quantum chemistry, but has received relatively little attention in quantum computation thus far.  From these considerations, we showed that in practical implementations of quantum phase estimation using a first order Suzuki-Trotter approximation, one expects a scaling cost on the order of $O(M^5 \log M)$ with respect to number of spin-orbitals, rather than more pessimistic estimates of $O(M^8)$-$O(M^9)$\cite{Wecker:2013,Hastings:2014} or $O(M^{5.5})$-$O(M^{6.5})$\cite{Poulin:2014}  related to the use of unphysical random integral distributions or the restriction to molecules too small to observe the effects of physical locality.  We believe that the combination of the algorithmic improvements suggested by Poulin and Hastings et al~\cite{Poulin:2014,Hastings:2014} with strategies that exploit locality presented here will result in even greater gains, and more work is needed in this area.

We also considered the cost of Hamiltonian averaging, an alternative to quantum phase estimation with minimal coherence time requirements beyond state preparation.  This method has some overhead with respect to quantum phase estimation, scaling as $O(M^6)$ in the number of spin-orbitals, but has significant practical advantages in coherence time costs, as well as the ability to make all measurements in parallel.  This method can at best give the energy of the state provided when oracle guesses are imperfect, however it can easily be combined with a variational or adiabatic approach to improve the accuracy of the energy estimate.  Moreover, while quantum phase estimation promises to be able to remove the bias of imperfect oracle guesses, we demonstrated how the cost of removal may strongly depend on how imperfect the guesses are.

Finally we analyzed the impact of locality on a complementary approach for quantum chemistry, namely adiabatic quantum computation.  This approach does not have a known direct dependence on the quality of guess states provided by an oracle, and can in fact act as the state oracle for the other approaches discussed here.

In all cases, the consideration of physical locality greatly improves the outlook for quantum chemistry on a quantum computer, and in light of the goal of quantum chemistry to study physical systems rather than abstract constructs, it is the correct to include this physical locality in any analysis pertaining to it.  We believe that with these and other developments made in the area of quantum computation, quantum chemistry remains one of the most promising applications for exceeding the capabilities of current classical computers.

\section{Acknowledgments}
J.M. is supported by the Department of Energy Computational Science Graduate Fellowship under grant number DE-FG02-97ER25308. P.J.L. acknowledges the National Science Foundation grant number PHY-0955518.  A.A-G. and P.J.L. appreciate the support of the Air Force Office of Scientific Research under Award No. FA9550-12-1-0046.  R.B. and A.A.-G. are grateful to the National Science Foundation for Award No. CHE-1152291.

\bibliographystyle{apsrev4-1}
\bibliography{QCQCPerspective,library}
\end{document}